%Paper: gr-qc/9405069
%From: "Dr. Beverly Berger" <berger@vela.acs.oakland.edu>
%Date: Fri, 27 May 1994 19:50:09 -0400

%revised version
\magnification1200
\baselineskip=24pt
\belowdisplayskip=24pt
{\nopagenumbers
\topglue 1in
{\parskip=0pt
\centerline{DETECTION OF COMPUTER GENERATED
GRAVITATIONAL}

\centerline{WAVES IN NUMERICAL COSMOLOGIES}}
\vskip 1in
\centerline{Beverly K. Berger$^a$, David Garfinkle$^a$, and Vijaya Swamy$^a$}
\vskip 1in
\centerline{$^a$Physics Department, Oakland University, Rochester, MI
48309}
\centerline{May 24, 1994}
This paper received Honorable Mention in the 1994 Gravity Essay
Contest.
\vskip .5in
\noindent
Running Title: Detection of Computer Generated Gravitation Waves

\vskip .5in
\noindent
%Send Correspondence to:  Beverly K. Berger, before 21 August: L-413,
%IGPP, Lawrence Livermore National Laboratory, P.O. Box 808, Livermore,
%CA  94551, (510)606-1681; after 4 September: Physics Department,
%Oakland University, Rochester, MI  48309, (313)370-3416; email:
%berger@vela.acs.oakland.edu
 \vfill
\eject

{\parindent=0pt ABSTRACT}

We propose to study the behavior of complicated numerical solutions to
Einstein's equations for generic cosmologies by following the geodesic
motion of a swarm of test particles.  As an example, we consider a
cylinder of test particles initially at rest in the plane symmetric Gowdy
universe on $T^3 \times R$.  For a circle of test particles in the
symmetry plane, the geodesic equations predict evolution of the circle
into distortions and rotations of an ellipse as well as motion
perpendicular to the plane.  The evolutionary sequence of ellipses
depends on the initial position of the circle of particles.  We display
snapshots of the evolution of the cylinder.
\vfill
\eject}

\pageno=1
{\parindent=0pt I.  Introduction.}

Modern computer technology allows numerical investigation of nontrivial
solutions to Einstein's equations.  In particular, a program of research is
now in progress to discover the nature of the Big Bang in generic
spatially inhomogeneous universes [1].  Yet this ability to tackle
complicated spacetimes brings with it the problem of how to interpret the
computer's output.  This problem is especially acute in the cosmological
context where the interpretational familiarity of an asymptotically flat
region is absent.

As an example, consider spatially inhomogeneous vacuum cosmologies
on $T^3 \times R$ that possess a single spatial $U(1)$ symmetry (the
$U(1)$ models).  According to Moncrief [2, 1], these models can be
described by a metric written in terms of five degrees of freedom at each
spatial point.  (These are equivalent to the two degrees of freedom that
describe the gravitational waves with the others restricted by constraints
and coordinate conditions.)  Clearly, graphical display of the amplitudes
of all the variables on the spacetime coordinate grid will not yield a
physical understanding of the model.  We propose as an alternative a
numerical gravitational wave detector:  the geodesic motion of a swarm
of test masses in the computer generated spacetime.

As an initial implementation of this idea, we present here snapshots from
computer movies of the geodesic motion of a cylinder of test masses in
the simpler Gowdy $T^3 \times R$ cosmology [3].  (Although we shall
continue to refer to a cylinder, $T^3$ topology means that the detector is
actually a 2-torus.)  The numerical methodology for solving Einstein's
equations for this model (described elsewhere [1]) allows it to be
regarded as a test case for the $U(1)$ models.  The Gowdy model can be
interpreted as plane symmetric gravitational waves propagating in one
direction in a background cosmology with the two polarizations of the
waves nonlinearly coupled.  Extensive numerical investigations [1, 4]
have shown that as Gowdy models evolve toward the singularity, the
nonlinear wave interactions cause the appearance of ever shorter
wavelength modes.  Depending on initial conditions, this nonlinear
growth gets suppressed as the model becomes velocity dominated [1, 5].
(This means that time derivatives dominate spatial derivatives in
Einstein's equations so that each point in the universe evolves
independently causing the spatial profile of the metric variables to
become frozen.)

For a particular class of initial conditions, we shall compare the behavior
of our ``detector'' in a universe with little growth of small scale spatial
structure to that in one with significant growth.  We shall consider both
``swarm of particles'' and ``rendered spatial volume'' visualizations of
the detector.  These, in turn, will be compared to the amplitudes of the
metric variables.  Directions for future enhancements of the detector will
also be discussed.

{\parindent=0pt II.  The Gowdy $T^3 \times R$ Model.}

The metric for this cosmology is
$$
d {s^2} = {e^{2 a}} \, \left ( - \, d {t^2} \; + \; d {\theta ^2} \right )
\; + \; t \, {e^P} \, \left [ d {\sigma ^2} \; + \; \left ( {e^{ - 2 P}} \, +
\, {Q^2} \right ) \, d {\delta ^2} \; + \; 2 Q \, d \sigma \, d \delta \right ]
\; \; \;
\eqno(1)$$
where $P$, $Q$, and $a$ depend only on $\theta$ and $t$. The two
Killing fields are
$ \left ( \partial / \partial \sigma \right ) $
and
$ \left ( \partial / \partial \delta \right ) $.  Einstein's equations for this
model consist of wave equations for $P$ and $Q$:
$$P,_{\tau \tau }-\;e^{-2\tau }P,_{\theta \theta }-\;e^{2P}\left( {Q,_\tau
^2-\;e^{-2\tau }Q,_\theta ^2} \right)=0,\eqno(2a)$$
$$Q,_{\tau \tau }-\;e^{-2\tau }Q,_{\theta \theta }+\;2\left( {P,_\tau
Q,_\tau
^{}-\;e^{-2\tau }P,_\theta Q,_\theta ^{}} \right)=0,\eqno(2b)$$
and the momentum and Hamiltonian constraints respectively for
$\lambda$ in terms of $P$ and $Q$:
$$\lambda ,_\theta -\;2(P,_\theta P,_\tau +\;e^{2P}Q,_\theta Q,_\tau
)=0,\eqno(3a)$$
$$\lambda ,_\tau -\;[P,_\tau ^2+\;e^{-2\tau }P,_\theta
^2+\;e^{2P}(Q,_\tau
^2+\;e^{-2\tau }\,Q,_\theta ^2)]=0,\eqno(3b)$$
where $\lambda = 4 a + \tau$ and $\tau = - \ln t$.

{\parindent=0pt III.  Geodesic equations.}

In order to construct our gravitational wave detector, we require the
geodesic equations for the Gowdy spacetime.  It is easily seen that
symmetries allow the detector to have a simple shape.
For $ u^a $
the tangent vector to a geodesic and $ \xi ^a $
a Killing vector, $ {u^a} {\xi _a} $
is a constant of the motion.  It then follows that the geodesics in the
Gowdy spacetime admit the following two constants of the motion:
$$
{c_1} = t \, {e^P} \, \left [ {{d \sigma } \over {d \rho }} \; + \; Q \;
{{d \delta } \over {d \rho }} \right ] \; \; \; ,
\eqno(4a)$$
$$
{c_2} = t \, {e^P} \, \left [ Q \; {{d \sigma } \over {d \rho }} \; +
\; \left ( {e^{- 2 P}} \, + \, {Q^2} \right ) \; {{d \delta } \over
{d \rho }} \right ] \; \; \; ,
\eqno(4b)$$
where
$ \rho $
is the affine parameter of the geodesic.  In particular,
geodesics for which
$ d \sigma / d \rho $
and
$ d \delta / d \rho $
vanish at the initial time, have
$ {c_1} = {c_2} = 0 $
so that
$ \sigma $
and
$ \delta $
are constant.  The motion of the particle is thus given by finding
$ \theta $
as a function of
$ t $.  This means that the detector can be constructed as a cylinder of test
masses with fixed coordinates in the $\sigma$-$\delta$ plane.

Note that, for the geodesics that describe the detector, the proper time
$ \Delta T $
is given by
$$
\Delta T = \int \, L \, d t
\eqno(5)$$
where the Lagrangian
$ L $
is given by
$$
L = {e^a} {{\left ( 1 \; - \; {{\left [ {{d \theta } \over {d t}}
\right ] }^2} \right ) }^{1/2}}  \; \; \; .
\eqno(6)$$
Since geodesics are those paths that extremize proper time, the geodesic
equation reduces to the Euler-Lagrange equations for the Lagrangian
$ L $,
$$
{{{d^2} \theta } \over {d {t^2}}} \; + \; \left (  1 \; - \; {{\left [
{{d \theta } \over {d t}} \right ] }^2} \right ) \, \left ( {{\partial a}
\over {\partial t}} \; {{d \theta } \over {d t}} \; + \; {{\partial a}
\over {\partial \theta }} \right ) = 0 \; \; \; .
\eqno(7)
$$
Note that, generically, each ring composing the cylindrical detector must
move in $\theta$ as the spacetime evolves.

In studying the approach to the singularity, it is helpful to use the
coordinate
$ \tau $ since the singularity occurs at $\tau = \infty$.
In terms of this time coordinate the equation of motion becomes
$$
{{{d^2} \theta } \over {d {\tau ^2}}} \; - \; {{d \theta } \over
{d \tau}} \; + \; \left ( {e^{2 \tau }} \; - \; {{\left [ {{d \theta }
\over {d \tau }} \right ] }^2} \right ) \, \left ( {e^{- 2 \tau }} \;
{{\partial a} \over {\partial \tau }} \; {{d \theta } \over {d \tau }} \; +
\; {{\partial a} \over {\partial \theta }} \right ) = 0 \; \; \; .
\eqno(8)$$
Our approach then is to integrate equation (7) or (8) for representative
rings of the cylindrical detector in the numerically generated background
spacetime (see Fig.~1).

{\parindent=0pt IV.  Motion of a ring of particles.}

Consider one of these rings with the given initial values of ($ \theta , t , d
\theta / d \rho , d t / d \rho $).
We wish to follow the shape of the ring as the particles
move.  The motion of the ring of test particles is described by a single
solution of equation (7); the particles which compose the ring differ from
each other only in the (constant) values of the coordinates
$ \sigma $
and
$ \delta $.
However, as the particles move they encounter changing metric functions
$ P $
and
$ Q $.
This gives rise to changing lengths associated with the constant spatial
coordinates
$ \sigma $
and
$ \delta $
and thus a changing shape of the ring. From equation (1),
the conformal metric of the symmetry plane is given by
$$
d {{\tilde s}^2} = {e^P} \, d {\sigma ^2} \; + \; \left ( {e^{- P}} \,
+ \, {e^P} {Q^2} \right ) \, d {\delta ^2} \; + \; 2 \, {e^P} \,
Q \, d \sigma \, d \delta \; \; \; .
\eqno(9)$$
Since all particles in the ring are described by the same
$ \theta ( t ) $,
distances between the particles are given by
$ {\sqrt t} d {\tilde s} $.
The overall size of the ring is then proportional to
$ \sqrt t $.
However, since we are only concerned with the shape of the ring, we will
consider only distances measured by the conformal metric
$ d {{\tilde s}^2} $.

We first treat the simpler case of a polarized Gowdy spacetime ($Q=0$)
where
$$
d {{\tilde s}^2} = {e^P} \, d {\sigma ^2} \; + \; {e^{- P}} \,
d {\delta ^2} \; \; \; .
\eqno(10)$$
In terms of
$ X \equiv {e^{P/2}} \sigma$ and $ Y \equiv {e^{- P / 2}} \, \delta $,
the conformal metric is flat.  Suppose that at the initial
time the ring has the shape of a unit circle:
$ {X_0 ^2} \, + \, {Y_0 ^2} = 1 $.
This means that
$$
{e^{P_0}} \, {\sigma ^2} \; + \; {e^{ - {P_0}}} \, {\delta ^2} = 1  \; \; \;
{}.
\eqno(11)
$$
Since the
($\sigma , \delta $)
coordinates of the particles are constant, equation (11)
holds at all times.  Expressing
($ \sigma , \delta $)
in terms of
($X, Y$),
we find
$$
{e^{{P_0} - P}} \, {X^2} \; + \; {e^{P - {P_0}}} \, {Y^2} = 1 \; \; \; ,
\eqno(12)$$
the equation for an ellipse.  Thus the ring of particles that
started out as a circle is distorted into an ellipse by the gravitational
waves.

We now consider the general case of an unpolarized Gowdy spacetime.
Define the
coordinates
$ \bar X $
and
$ \bar Y $
by
$ {\bar X} \equiv {e^{P/2}} \, (\sigma + Q \delta ) , \; {\bar Y}
\equiv {e^{- P / 2}} \, \delta $ yielding
$ d {{\tilde s}^2} = d {{\bar X}^2} \, + \, d {{\bar Y}^2} $.
However, the
$ \bar X $
axis rotates relative to a parallel propagated frame.  Let
$ \psi $
be the angle of rotation and define
($X, Y$)
by
$$
{\bar X} = X \, \cos \psi \; + \; Y \, \sin \psi \; \; \; ,
\eqno(13a)$$
$$
{\bar Y} = Y \, \cos \psi \; - \; X \, \sin \psi \; \; \; .
\eqno(13b)$$
Then
$ d {{\tilde s}^2} = d {X^2} \, + \, d {Y^2} $
and the
$ X $
axis is nonrotating.  Thus expressing the positions of the particles in the
ring in terms of the
($X, Y$)
coordinate system will tell us how much the ring is rotated as well as
how much it is distorted.

We now calculate the rotation angle
$ \psi $.
Let
($ {n_a} , {m_a} $)
be an orthonormal basis in the
($ {\bar X} , {\bar Y}$)
directions (described e.g.~in the nonrotating coordinate system).  Let
$ u^a $
be the tangent vector to the geodesic at the center of the ring.  Then the
changes in the basis vectors are related by
$$
{u^a} {\nabla _a} {n_b} = \left ( {u^a} {\nabla _a} \psi \right ) \,
{m_b} \; \; \; .
\eqno(14)$$
For
$ {\xi ^a} \equiv {{( \partial / \partial \sigma )}^a} $, the metric and the
definition of $(\bar X, \bar Y)$ lead to
$$
{n_a} = {t^{- 1 / 2}} \, {e^{- P / 2}} \, {\xi _a} \; \; \; ,
\eqno(15)$$
$$
{\xi _a} = t \, {e^P} \, \left ( {\partial _a} \sigma \, + \,
Q {\partial _a} \delta \right ) \; \; \; .
\eqno(16)$$
Using the fact that
$ \xi ^a $
is a Killing vector,  we find
$$
{u^a} {\nabla _a} {n_b} = {\xi _b} \, {u^a} {\partial _a} \left (
{t^{- 1 / 2}} \, {e^{- P / 2}} \right ) \; + \;  {t^{- 1 / 2}} \,
{e^{- P / 2}} \, {u^a}
{\partial _{[a}} {\xi _{b ]}} \; \; \; .
\eqno(17)$$
Some straightforward algebra yields
$$
{u^a} {\nabla _a} {n_b} = {1 \over 2} \; {t^{1/2}} \, {e^{P / 2}} \,
\left ( {u^a} {\nabla _a} Q \right ) \, {\partial _b} \delta
$$
$$
= {1 \over 2} \; {e^P} \, \left ( {u^a} {\nabla _a} Q \right ) \,
{m_b} \; \; \; .
\eqno(18)$$
We therefore have
$$
{u^a} {\nabla _a} \psi = {1 \over 2} \; {e^P} \, {u^a} {\nabla _a} Q
$$
and thus
$$
\psi = {1 \over 2} \; \int \; {e^P} \; \left ( {{\partial Q} \over
{\partial t}} \; + \; {{\partial Q} \over {\partial \theta }} \;
{{d \theta } \over {d t}} \right ) \; d t \; \; \; .
\eqno(19)$$

We can now find the shape of the ring of particles.  Since the ring
starts out as a unit circle, we have
$$
1 = {{\bar X}_0 ^2} \; + \; {{\bar Y}_0 ^2} = {e^{P_0}} \, {{\left (
\sigma \, + \, {Q_0} \delta \right ) }^2} \; + \; {e^{- {P_0}}} \,
{\delta ^2} \; \; \; .
\eqno(20)$$
Since
$ \sigma $
and
$ \delta $
are constant, equation (20) holds at all times.  Expressing
$ \sigma $
and
$ \delta $
in terms of
$ \bar X $
and
$ \bar Y $
gives
$$
{e^{{P_0} - P}} \, {{\left [ {\bar X} \, + \, \left ( {Q_0} \, - \,
Q \right ) \, {e^P} \, {\bar Y} \right ] }^2} \; + \; {e^{P - {P_0}}} \,
{{\bar Y}^2} = 1 \; \; \; .
\eqno(21)$$
This is again the equation of an ellipse.  Finding the angle
$ \psi $
then allows us to rotate this ellipse to the
($X, Y$)
coordinate system.

Our explicit procedure is to construct an initial set of unit circle rings
uniformly distributed along the $\theta$ axis with zero initial velocity in
$\theta$.  A previously computed numerical spacetime $\{P, Q, a \}
(\theta,\tau)$ is used in the geodesic equation (8) to evolve $\theta (\tau)$
for each ring.  When preselected values of $\tau$ are reached, the shape
of each ring is computed using equation (21) with equations (13) to
construct the ellipse in a nonrotating frame.  For example, expressing
$(X,Y)$ in terms of polar coordinates $(r, \zeta)$ in the
$\sigma$-$\delta$
plane gives $r(\zeta)$ for the ellipse directly from equation (21).

{\parindent=0pt V.  Results.}

Various techniques have been tested to visualize the evolution of the
detector.  Two of these are illustrated in the figures.  First we consider
explicit display of the detector's component rings.  (The visualization
here was constructed using D$^2$ Software MacSpin 3.0 on a
Macintosh Quadra 660AV.  A similar construction using custom
software written by one of us (VS) runs under X-Windows on a Unix
workstation.)  Alternatively, we can display a rendered volume of the
entire cylinder with interpolations to find values in between the selected
rings.  The displayed values are defined as the region inside each ring.
The location of the ellipse in a $4 \times 4$ square was found by
transforming each point of an array on the square to the coordinate
system defined by the major and minor axes of the ellipse.  (This
visualization uses Spyglass Dicer 3.0 on the Macintosh.  One of us (VS)
has written custom software for visualization under X-Windows.)

The detector at $\tau = 0$ is shown in Fig.~1 as a unit cylinder.  Figure 2
illustrates the distortion of the ellipses for a polarized
Gowdy background later in the evolution.  Note
that, as expected, there is no rotation of the ellipse.
Agreement with equation (11) has been checked.

More interesting is a comparison between two Gowdy universes which
differ only in the value of a parameter $v_0$ in their initial conditions.
This parameter controls the nonlinear growth of small scale spatial
structure in these models [1, 4].  In Figs.~3a and 3b, the complete
evolution of
$P$, $Q$, and $\lambda$ is displayed for $v_0 = 0.5$ and $v_0=5$
respectively.  Note the increased spatial structure in $P$ for the larger
$v_0$.  (The interpretation of these features is discussed elsewhere
[1, 4].)  Figures 4 and 5 display snapshots of the evolution of the detector
(using both visualizations) at three values of $\tau$ for $v_0 = 0.5$ and
$v_0=5$ respectively.  In all cases, the values of the metric functions
$P$ and $Q$ (which determine the detector's shape in the
$\sigma$-$\delta$ plane)
are also displayed.  Within the limited resolution of the
detector as illustrated here (due to convenience rather than principle), the
features of the spacetime are clearly seen.  Note both the spatial variation
in orientations of the ellipses and the rotation (as well as distortions) of
the ellipses with time.

{\parindent=0pt V.  Discussion.}

We have demonstrated the usefulness of a numerical gravitational wave
detector in the visualization of spatially inhomogeneous cosmologies.
For application to the $U(1)$ models, the absence of the symmetry plane
probably means that the detector should consist of a swarm of
individually followed test particles.  The initial conditions of the swarm
could be adjusted to explore a variety of features of the $U(1)$ spacetime
on a variety of spatial scales.

{\parindent=0pt  Acknowledgments.}

BKB was supported in part by National Science Foundation Grant
PHY93-0559.  Computations were performed using the facilities of the
National Center for Supercomputing Applications at the University of
Illinois.  BKB and DG wish to thank the Astronomy Department of the
University of Michigan for hospitality. VS's work was performed in
partial fulfillment of the requirements for the MS in Physics at Oakland
University.

\vfill
\eject

{\parindent=0pt REFERENCES

[1]  Berger, B.~K. and Moncrief, V. (1993)  {\it Phys. Rev. D} {\bf
48}, 4676--4687.

[2]  Moncrief, V. (1986)  {\it Ann. Phys. (N.Y.)} {\bf 167}, 118.

[3]   Gowdy, R.~H. (1971)  {\it Phys. Rev. Lett.} {\bf 27}, 826,
E1102; Gowdy, R.~H. (1974).  {\it Ann. Phys. (N.Y.)} {\bf 83}, 203.

[4]  Berger, B.~K., Garfinkle, D., Moncrief, V., and Swift, C.~M.
(1994)  {\it Proceedings of the NATO Advanced Research Workshop on
Deterministic Chaos in General Relativity} ed. by D. Hobill et al
(Plenum);  Berger, B.~K., Garfinkle, D., Grubi\v si\'c, B., Moncrief,
V., and Swamy, V. (1994)  {\it Proceedings of the Lanzcos
Symposium}.

[5]  Grubi\v si\'c, B. and Moncrief, V. (1993) {\it Phys.~Rev.~D} {\bf
47}, 2371.}
\vfill
\eject
{\parindent=0pt
\parskip=18pt FIGURE CAPTIONS

Fig.~1.  The initial state of the detector.  (a)  The volume of space within
the unit rings in the $\sigma$-$\delta$ plane that compose the detector.
The $\sigma$-$\delta$ plane is represented as a $25 \times 25$ array of
points for $-2 \le X,Y \le 2$. (See Section IV.)
(b)  Each of 10 rings that constitutes the cylinder is displayed as 100
points at fixed angle $\zeta$ in the $\sigma$-$\delta$ plane.
In each case, 10 frames of the evolution are saved.

Fig.~2.  The detector in a polarized model.  A polarized model with
$P(\theta,0) = 0$, $P,_\tau(\theta,0) = 0.5 \cos \theta$, $Q(\theta,0) = 0$,
and $Q,_\tau(\theta,0) = 0$ is evolved up to $\tau = 9.2$.  The shape of
the detector at $\tau = 5.2$ is displayed for both visualizations.

Fig.~3.  The unpolarized background spacetimes.  Unpolarized models
with $P(\theta,0) = 0$, $P,_\tau(\theta,0) = v_0 \cos \theta$, $Q(\theta,0)
= \cos \theta$, and $Q,_\tau(\theta,0) = 0$ are evolved up to $\tau_f =
8.40$.  The complete evolution of the metric functions $P$, $Q$, and
$\lambda$ is shown for (a) $v_0 = 0.5$ and (b) $v_0 = 5$.  In all cases,
$0 \le \theta \le 2 \pi$ and $0 \le \tau \le \tau_f$.  The vertical scale
indicates the value of the appropriate metric variable.

Fig.~4.  Snapshots of the detector's evolution in the unpolarized model
with $v_0 = 0.5$.  Both visualizations of the detector and slices in time
of $P$ and $Q$ vs. $\theta$ are shown for (a) $\tau = 2.80$, (b) $\tau =
4.67$, and (c) $\tau = 6.53$.  For the metric variables, the horizontal
scale is $0 \le \theta \le 2 \pi$ while the vertical scale is the same as that
in
Fig.~3a.

Fig.~5.  The same as Fig.~4 for $v_0 = 5$.  The vertical scale for the
metric variables is from Fig.~3b. Here 20 rings were used to describe the
detector.}

\end